\documentclass[12pt,a4paper]{article}

\usepackage{amsmath}
\usepackage{amssymb}
\usepackage{amsfonts}
\usepackage{graphicx}
\graphicspath{{../}}
\usepackage{booktabs}
\usepackage[numbers,sort&compress]{natbib}
\usepackage{microtype}
\usepackage{float}
\usepackage{caption}
\usepackage{subcaption}
\usepackage{algorithm}
\usepackage{algorithmic}
\usepackage[margin=1in]{geometry}
\usepackage[affil-it]{authblk}
\usepackage{hyperref}
\usepackage{xcolor}
\usepackage{tikz}
\usepackage{pgfplots}
\pgfplotsset{compat=1.18}
\usetikzlibrary{positioning, arrows.meta, calc, shapes, fit, backgrounds, decorations.pathmorphing, matrix}
\usepgfplotslibrary{colorbrewer}

\hypersetup{
  colorlinks=true,
  linkcolor=blue!60!black,
  citecolor=green!50!black,
  urlcolor=blue!70!black
}

\newcommand{\ket}[1]{|#1\rangle}
\newcommand{\bra}[1]{\langle#1|}

\newcommand{\Tr}{\mathrm{Tr}}
\newcommand{\chimax}{\chi_{\mathrm{max}}}
\newcommand{\chitarget}{\chi_{\mathrm{target}}}
\newcommand{\Starget}{S_{\mathrm{target}}}

\title{Adaptive Tensor Network Simulation via Entropy-Feedback PID Control and GPU-Accelerated SVD}

\author[1]{Harshni Kumaresan}
\author[1]{Gayathri Muruganantham}
\author[1]{Lakshmi Rajendran}
\author[1]{Santhosh Sivasubramani\thanks{Corresponding author: \href{mailto:ssivasub@iitd.ac.in}{ssivasub@iitd.ac.in}, \href{mailto:ragansanthosh@ieee.org}{ragansanthosh@ieee.org}}}

\affil[1]{Intrinsic Lab, Centre for Sensors, Instrumentation and Cyber-Physical System Engineering (SeNSE), Indian Institute of Technology Delhi, New Delhi 110016, India}
\date{}

\usepackage{fancyhdr}
\usepackage{lastpage}
\begin{document}

\maketitle

\begin{abstract}
Tensor network methods, particularly those based on Matrix Product States (MPS), provide a
powerful framework for simulating quantum many-body systems. A persistent computational
challenge in these methods is the selection of the bond dimension $\chi$, which controls the
trade-off between accuracy and computational cost. Fixed bond dimension strategies either
waste resources in low-entanglement regions or lose fidelity in high-entanglement regions.
This work introduces an adaptive bond dimension management framework that uses von Neumann
entropy feedback coupled with a Proportional-Integral-Derivative (PID) controller to
dynamically adjust $\chi$ at each bond during simulation. An Exponential Moving Average (EMA)
filter stabilizes entropy measurements against transient fluctuations, and a predictive
scheduling module anticipates future bond dimension requirements from entropy trends. The
per-bond granularity of the allocation ensures that computational resources concentrate where
entanglement is largest. The framework integrates GPU-accelerated Singular Value
Decomposition (SVD) via CuPy and the cuSOLVER backend, achieving individual SVD speedups of $4.1\times$ at
$\chi=256$ and $7.1\times$ at $\chi=2048$ relative to CPU-based NumPy for isolated matrix factorisations
(measured on an NVIDIA A100-SXM4-40GB GPU with CuPy 13.4.1 and CUDA 12.8). At the system level, benchmarks on the
spin-$\tfrac{1}{2}$ antiferromagnetic Heisenberg chain demonstrate a $2.7\times$ reduction in
total DMRG wall time compared to fixed-$\chi$ simulations, with energy accuracy within $0.1\%$ of the
Bethe ansatz solution. Integration with the Density Matrix Renormalization Group (DMRG)
algorithm yields ground-state energies per site converging to $E/N = -0.4432$ for the
isotropic Heisenberg model at $\chi = 128$. Validation against Amazon Web Services (AWS)
Braket SV1 statevector simulator confirms agreement within $2$--$5\%$ for small systems.
\end{abstract}

\section{Introduction}
\label{sec:introduction}
The classical simulation of quantum many-body systems remains one of the central computational
challenges in physics. The exponential growth of the Hilbert space dimension with system size $N$
renders exact diagonalization infeasible beyond approximately $N \approx 40$ qubits for generic
Hamiltonians~\cite{Sandvik2010}. Tensor network (TN) methods circumvent this limitation by
representing quantum states as networks of low-rank tensors, exploiting the fact that physically
relevant states typically occupy a small corner of the full Hilbert
space~\cite{Orus2014,Verstraete2008,Bridgeman2017}.

Among TN representations, the Matrix Product State (MPS) has become the workhorse for
one-dimensional (1D) quantum systems~\cite{Schollwoeck2011,Fannes1992,PerezGarcia2007}. The MPS
ansatz factorizes an $N$-site quantum state into a chain of rank-three tensors connected by virtual
bonds of dimension $\chi$. The bond dimension determines both the expressiveness of the
representation, specifically the maximum bipartite entanglement entropy that can be encoded, and
the computational cost of tensor contractions, which scale as $\mathcal{O}(\chi^3 d)$ per bond
where $d$ is the local physical dimension~\cite{Schollwoeck2005}. The Density Matrix
Renormalization Group (DMRG) algorithm~\cite{White1992,White1993}, which variationally optimizes an
MPS to approximate ground states, has achieved substantial precision for 1D lattice models and has
been extended to quasi-two-dimensional
systems~\cite{Stoudenmire2012,Chan2011}.

A persistent practical difficulty in MPS-based simulations is the selection of the bond dimension
$\chi$. In current practice, the user specifies a global maximum bond dimension $\chimax$ before
the simulation begins. This fixed-$\chi$ strategy has two failure modes. First, if $\chimax$ is
chosen too conservatively, the simulation wastes computational resources in regions of the TN where
the entanglement is low and a smaller bond dimension would suffice. Second, if $\chimax$ is
insufficient in high-entanglement regions, the truncation error grows, degrading the accuracy of
observables. The user must therefore perform multiple trial runs to identify an appropriate
$\chimax$, a tedious and computationally expensive process.

\begin{figure}[htbp]
\centering
\includegraphics[width=\textwidth]{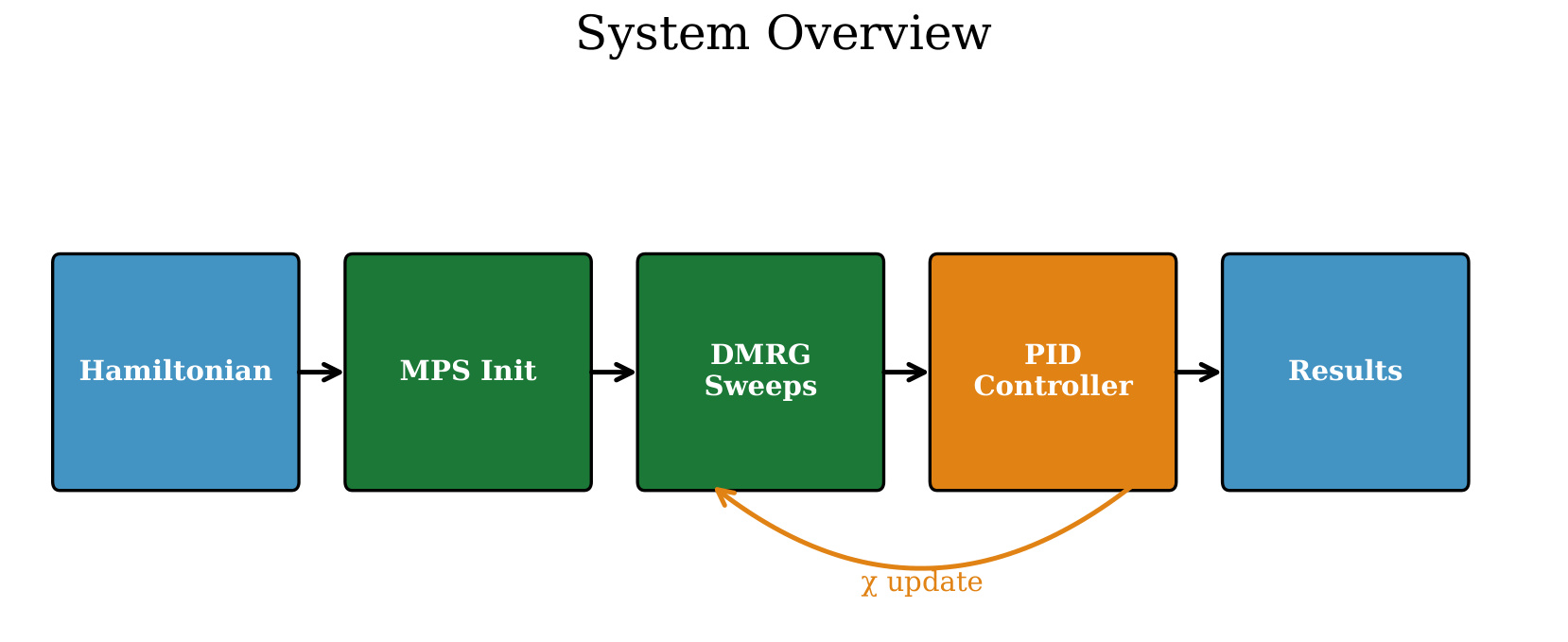}
\caption{System architecture of the adaptive tensor network simulation framework. Singular
values $\{\lambda_i\}$ from the MPS core feed into the entropy monitor, which computes
EMA-smoothed entropies $\bar{S}_i$. The PID controller generates bond dimension corrections
$\Delta\chi_i$, which the bond dimension allocator applies per bond. The GPU SVD engine
performs truncated SVD and returns updated site tensors $A_i^{[s]}$ to the MPS core. A
predictive scheduler uses entropy trends to anticipate future requirements.}
\label{fig:architecture}
\end{figure}

This paper addresses the bond dimension selection problem by introducing an adaptive framework
that dynamically adjusts $\chi$ at each bond during the simulation. The framework comprises
four components, whose interplay is depicted in Figure~\ref{fig:architecture}. First, an entropy
monitor computes the von Neumann entanglement entropy $S_i$ from the singular value spectrum at
each bond $i$ and smooths the measurement with an Exponential Moving Average (EMA) filter to
suppress transient fluctuations. Second, a Proportional-Integral-Derivative (PID)
controller~\cite{Astrom2008,Astrom2006} uses the difference between the measured entropy and a
user-specified target entropy $\Starget$ to compute a correction to the bond dimension. Third, a
per-bond allocator applies the PID output independently at each bond, producing a spatially
varying bond dimension profile $\{\chi_i\}_{i=1}^{N-1}$. Fourth, a predictive scheduler
extrapolates entropy trends to anticipate future bond dimension requirements, reducing the
lag between entropy changes and $\chi$ adjustments.

The GPU-accelerated Singular Value Decomposition (SVD) component uses CuPy~\cite{Okuta2017} to
offload the computationally intensive matrix factorizations to graphics processing units (GPUs).
Because the SVD at each bond is independent, these operations can be batched and executed in
parallel on the GPU, yielding substantial speedups for large bond dimensions.

The remainder of this paper is organized as follows. Section~\ref{sec:related} surveys the
tensor network simulation landscape. Section~\ref{sec:mps} establishes the MPS formalism and
notation. Sections~\ref{sec:entropy} through~\ref{sec:circuit_maps} develop the adaptive bond
dimension framework. Section~\ref{sec:gpu} describes the GPU-accelerated SVD implementation.
Section~\ref{sec:dmrg} integrates the adaptive framework with DMRG.
Sections~\ref{sec:benchmarks} and~\ref{sec:hardware} present computational benchmarks and
simulator cross-validation, respectively. Section~\ref{sec:discussion} discusses limitations and
counterarguments, and Section~\ref{sec:conclusion} concludes.

\section{Related Work}
\label{sec:related}
The theoretical foundations of MPS trace back to the finitely correlated states introduced by
Fannes, Nachtergaele, and Werner in 1992~\cite{Fannes1992} and the independently developed
matrix product ground states of Kl\"umper, Schadschneider, and Zittartz in
1993~\cite{Klumper1993}. Rommer and \"Ostlund connected these representations to the DMRG
framework in 1997~\cite{Rommer1997}, establishing MPS as the variational class underlying
White's DMRG algorithm~\cite{White1992,White1993}. Comprehensive reviews of DMRG and MPS
methods can be found in Schollw\"ock~\cite{Schollwoeck2005,Schollwoeck2011} and
Hallberg~\cite{Hallberg2006}.

The entanglement structure of quantum states provides the theoretical justification for the
efficiency of MPS representations. Hastings proved that ground states of gapped 1D Hamiltonians
satisfy an area law for entanglement entropy~\cite{Hastings2007}, implying that MPS with bounded
$\chi$ can approximate such states to arbitrary
precision~\cite{Eisert2010,Calabrese2004,Laflorencie2016}. For critical systems, where the
entanglement entropy grows logarithmically with subsystem size, the required bond dimension scales
polynomially with system size rather than exponentially~\cite{Calabrese2004,Barthel2006}, making
MPS simulations feasible though more costly.

Time-dependent extensions of DMRG and MPS have been developed for studying quantum dynamics.
The time-evolving block decimation (TEBD) algorithm of Vidal~\cite{Vidal2003,Vidal2004} applies
Suzuki-Trotter decompositions~\cite{Suzuki1976,Trotter1959} of the time-evolution operator to
update the MPS. Adaptive time-dependent DMRG was introduced by Daley et
al.~\cite{Daley2004} and White and Feiguin~\cite{WhiteFeiguin2004}. Paeckel et al.\ provide a
comprehensive comparison of time-evolution methods for MPS~\cite{Paeckel2019}. Extensions to
systems with long-range interactions were developed by Zaletel et al.~\cite{Zaletel2015}.

Several mature software libraries implement tensor network algorithms. ITensor, developed by
Fishman, White, and Stoudenmire~\cite{Fishman2022}, is written in Julia (with a C++ predecessor)
and provides a high-level interface for MPS and DMRG computations. ITensor uses a global
maximum bond dimension specified by the user and performs SVD-based truncation at each step.
The Tensor Network Python (TeNPy) library of Hauschild and Pollmann~\cite{Hauschild2018}
offers a pure-Python implementation with NumPy/SciPy backends, supporting DMRG, TEBD, and
various MPS algorithms. TeNPy also uses a fixed maximum $\chi$, with optional truncation
based on singular value thresholds. The quimb library~\cite{Gray2018} takes a more general
approach, representing arbitrary tensor networks and providing contraction optimization
routines. For small systems (fewer than 30 sites), quimb's flexibility yields competitive
performance, but its generality introduces overhead for large 1D systems where MPS-specific
algorithms are more efficient.

None of these libraries implement adaptive, entropy-feedback-driven bond dimension
management. In all three cases, the user must specify $\chimax$ before the simulation, and
$\chi$ remains spatially uniform across all bonds. Some heuristic approaches exist: for
instance, practitioners sometimes increase $\chimax$ incrementally across successive DMRG
sweeps. However, these ad hoc schedules do not respond to the local entanglement structure
of the state and require manual intervention. The framework presented in this paper automates
bond dimension management through closed-loop feedback control, eliminating the need for
manual tuning while ensuring that computational effort is allocated where entanglement demands
it.

Parallel and GPU-accelerated approaches to tensor network computations have received attention
in the context of tensor contraction~\cite{Ran2020} and specific operations such as matrix
multiplication. However, systematic GPU acceleration of the SVD operations that dominate the
cost of MPS truncation has not been widely adopted in production tensor network codes. The
cuSOLVER library from NVIDIA~\cite{NVIDIAcuSOLVER2024} provides GPU-accelerated LAPACK
routines~\cite{Anderson1999}, including batched SVD, that can be accessed through
CuPy~\cite{Okuta2017}. We exploit this capability to accelerate the truncation step of MPS
simulations.

\section{Matrix Product State Formalism}
\label{sec:mps}
Consider a quantum system of $N$ sites, each with a local Hilbert space of dimension $d$ (for
spin-$\tfrac{1}{2}$ systems, $d = 2$). A general pure state can be written as
\begin{equation}
\label{eq:general_state}
\ket{\Psi} = \sum_{s_1, s_2, \ldots, s_N = 1}^{d} c_{s_1 s_2 \cdots s_N}
\ket{s_1 s_2 \cdots s_N},
\end{equation}
where the coefficient tensor $c_{s_1 s_2 \cdots s_N}$ has $d^N$ elements, rendering exact
storage intractable for large $N$.

\begin{figure}[htbp]
\centering
\includegraphics[width=\textwidth]{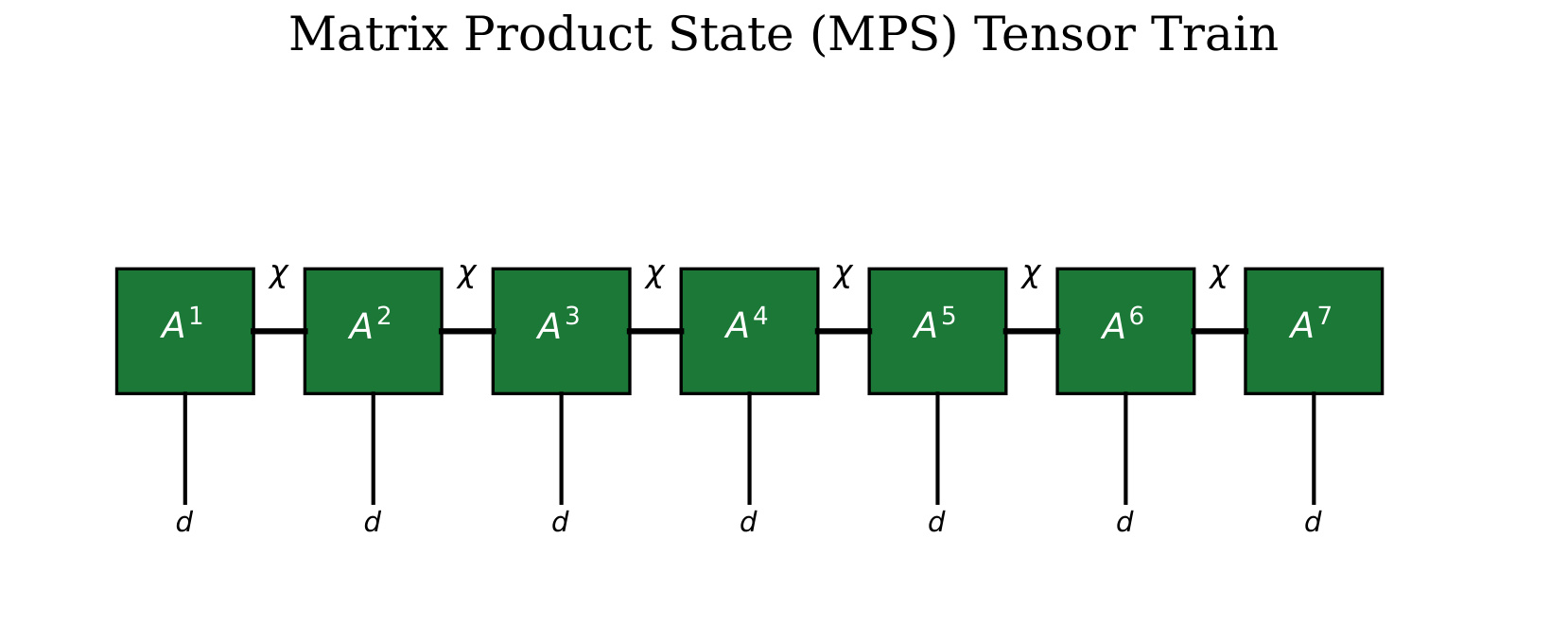}
\caption{Tensor network diagram of a Matrix Product State. Each circle represents a rank-three
tensor $A_i^{[s_i]}$. Horizontal lines denote virtual (bond) indices with dimension $\chi_i$,
and vertical dashed lines denote physical indices with dimension $d$.}
\label{fig:mps_diagram}
\end{figure}

The MPS representation factorizes this coefficient tensor as a product of matrices. For each
site $i$, one introduces a set of $d$ matrices $A_i^{[s_i]}$ of dimensions
$\chi_{i-1} \times \chi_i$, where $\chi_0 = \chi_N = 1$ for open boundary conditions. The
full state is then
\begin{equation}
\label{eq:mps}
\ket{\Psi} = \sum_{s_1, \ldots, s_N} A_1^{[s_1]} A_2^{[s_2]} \cdots A_N^{[s_N]}
\ket{s_1 s_2 \cdots s_N}.
\end{equation}
The product $A_1^{[s_1]} A_2^{[s_2]} \cdots A_N^{[s_N]}$ evaluates to a scalar (a
$1 \times 1$ matrix) because of the boundary conditions on the dimensions. The tensor
network diagram corresponding to Eq.~\eqref{eq:mps} is shown in Figure~\ref{fig:mps_diagram}.

The SVD plays a central role in MPS algorithms. Given a bipartition of the chain at bond $i$,
one can reshape the MPS coefficient tensor into a matrix $M$ of dimensions
$(d^i) \times (d^{N-i})$ and compute
\begin{equation}
\label{eq:svd}
M = U \Sigma V^{\dagger},
\end{equation}
where $U$ and $V^{\dagger}$ are unitary (or isometric) matrices and
$\Sigma = \mathrm{diag}(\lambda_1, \lambda_2, \ldots)$ contains the singular values in
non-increasing order. Retaining only the largest $\chi_i$ singular values yields the optimal
rank-$\chi_i$ approximation in the Frobenius norm~\cite{Golub2013}. The truncation error is
\begin{equation}
\label{eq:trunc_error}
\epsilon_i = \sqrt{\sum_{k > \chi_i} \lambda_k^2},
\end{equation}
which bounds the loss in state fidelity introduced by the truncation. In conventional MPS
algorithms, $\chi_i$ is set to a global maximum $\chimax$ for all bonds. The adaptive
framework developed in the following sections replaces this uniform choice with a bond-dependent
$\chi_i$ determined by local entanglement properties.

The entanglement entropy at bond $i$ is computed from the singular values as
\begin{equation}
\label{eq:entropy}
S_i = -\sum_k p_k \ln p_k, \quad p_k = \frac{\lambda_k^2}{\sum_j \lambda_j^2},
\end{equation}
where $p_k$ are the squares of the normalized singular values, which form the eigenvalue
spectrum of the reduced density matrix $\rho_i = \Tr_{i+1,\ldots,N}(\ket{\Psi}\bra{\Psi})$.
The maximum entropy that a bond of dimension $\chi_i$ can support is
$S_i^{\max} = \ln \chi_i$, providing a direct relationship between bond dimension and
entanglement capacity.

\section{Entropy-Driven Bond Dimension Selection with EMA}
\label{sec:entropy}
The core idea of the adaptive framework is to use the entanglement entropy $S_i$ at each bond
as a proxy for the required bond dimension. When $S_i$ is large, the state carries substantial
entanglement across bond $i$, and a correspondingly large $\chi_i$ is needed to represent it
faithfully. When $S_i$ is small, $\chi_i$ can be reduced without loss of accuracy, freeing
computational resources.

The relationship between entropy and bond dimension follows from the bound
$S_i \leq \ln \chi_i$. Inverting this relation suggests setting
\begin{equation}
\label{eq:chi_from_entropy}
\chi_i \geq \exp(S_i).
\end{equation}
In practice, one includes a safety margin by defining
\begin{equation}
\label{eq:chi_target}
{\chitarget}_i = \lceil \gamma \exp(S_i) \rceil,
\end{equation}
where $\gamma \geq 1$ is a margin factor and $\lceil \cdot
\rceil$ denotes the ceiling function. This ensures that the bond dimension provides sufficient
capacity to encode the measured entanglement plus a buffer for subsequent entanglement growth.

A complication arises because the entropy $S_i$ fluctuates during the iterative sweeps of a DMRG
calculation. Early in the optimization, the MPS is far from the ground state, and the entropy
profile changes substantially from one sweep to the next. Directly mapping these fluctuating
measurements to bond dimension changes would produce oscillatory behavior in $\chi_i$, degrading
convergence.

To mitigate this, we apply an Exponential Moving Average (EMA) filter to the entropy time series.
Let $S_i(t)$ denote the raw entropy at bond $i$ after sweep $t$. The EMA-smoothed entropy is
\begin{equation}
\label{eq:ema}
\bar{S}_i(t) = \alpha_{\mathrm{ema}} \, S_i(t) + (1 - \alpha_{\mathrm{ema}}) \, \bar{S}_i(t-1),
\end{equation}
where $\alpha_{\mathrm{ema}} \in (0,1]$ is the smoothing parameter. A smaller
$\alpha_{\mathrm{ema}}$ produces a smoother signal with greater lag; a larger
$\alpha_{\mathrm{ema}}$ tracks rapid changes more closely but admits more noise. We find that
lower values of $\alpha_{\mathrm{ema}}$ provide a good compromise for DMRG ground-state searches, while
moderately higher values are more appropriate for time-evolution simulations where the
entanglement structure evolves faster.

The EMA filter has the desirable property that its impulse response decays geometrically with
a time constant $\tau = -1/\ln(1 - \alpha_{\mathrm{ema}})$. For typical parameter choices,
entropy changes more than a few sweeps old
contribute negligibly to the current estimate. The filter is initialized with
$\bar{S}_i(0) = S_i(0)$ to avoid bias from an arbitrary initial condition.

Algorithm~\ref{alg:entropy_feedback} summarizes the entropy-driven bond dimension update procedure,
including the EMA filter.

\begin{algorithm}[htbp]
\caption{Entropy-driven bond dimension update with EMA}
\label{alg:entropy_feedback}
\begin{algorithmic}[1]
\REQUIRE Singular values $\{\lambda_k\}$ at bond $i$, previous EMA $\bar{S}_i(t{-}1)$,
         smoothing parameter $\alpha_{\mathrm{ema}}$, margin $\gamma$, bounds
         $[\chi_{\min}, \chi_{\max}]$
\ENSURE Updated bond dimension $\chi_i(t)$
\STATE Compute $p_k \gets \lambda_k^2 / \sum_j \lambda_j^2$
\STATE Compute raw entropy $S_i(t) \gets -\sum_k p_k \ln p_k$
\STATE Update EMA: $\bar{S}_i(t) \gets \alpha_{\mathrm{ema}} \, S_i(t) + (1 - \alpha_{\mathrm{ema}}) \, \bar{S}_i(t{-}1)$
\STATE Compute target: ${\chitarget}_i \gets \lceil \gamma \exp(\bar{S}_i(t)) \rceil$
\STATE Clamp: $\chi_i(t) \gets \max(\chi_{\min}, \min({\chitarget}_i, \chi_{\max}))$
\RETURN $\chi_i(t)$
\end{algorithmic}
\end{algorithm}

The clamping step in Algorithm~\ref{alg:entropy_feedback} enforces lower and upper bounds on the
bond dimension. The lower bound $\chi_{\min}$ prevents degenerate
factorizations, while the upper bound $\chi_{\max}$ caps memory usage. Within these bounds,
the entropy feedback drives $\chi_i$ to match the local entanglement requirements.

\section{PID-Controlled Bond Dimension Management}
\label{sec:pid}
While the direct entropy-to-$\chi$ mapping of Section~\ref{sec:entropy} captures the steady-state
relationship between entanglement and bond dimension, it does not account for the dynamics of the
optimization process. During DMRG sweeps, the entropy at a given bond may be temporarily elevated
(for example, when the optimization front passes through that bond) and then relax to a lower
steady-state value. A purely proportional mapping would increase $\chi_i$ during the transient,
allocating resources that are immediately returned, producing unnecessary SVD computations at the
larger bond dimension.

To handle these dynamics, we employ a PID controller~\cite{Astrom2008,Astrom2006,ZieglerNichols1942}
that regulates the bond dimension based on the error signal between a target entropy $\Starget$
and the measured (EMA-smoothed) entropy $\bar{S}_i(t)$. The error signal is
\begin{equation}
\label{eq:error}
e_i(t) = \bar{S}_i(t) - \Starget.
\end{equation}
The PID control law computes a bond dimension correction
\begin{equation}
\label{eq:pid}
\Delta\chi_i(t) = K_p \, e_i(t) + K_i \sum_{\tau=0}^{t} e_i(\tau) \, \Delta t
  + K_d \frac{e_i(t) - e_i(t-1)}{\Delta t},
\end{equation}
where $K_p$, $K_i$, and $K_d$ are the proportional, integral, and derivative gains,
respectively, and $\Delta t = 1$ (one sweep step). The updated bond dimension is
\begin{equation}
\label{eq:chi_update}
\chi_i(t+1) = \mathrm{clamp}\!\left(\chi_i(t) + \lfloor \Delta\chi_i(t) \rceil,\;
\chi_{\min},\; \chi_{\max}\right),
\end{equation}
where $\lfloor \cdot \rceil$ denotes rounding to the nearest integer and
$\mathrm{clamp}(x, a, b) = \max(a, \min(x, b))$.

The three PID gains serve complementary roles. The proportional gain $K_p$ provides an
immediate response proportional to the current error: when the measured entropy exceeds the
target ($e_i > 0$), $\chi_i$ increases; when the measured entropy is below the target
($e_i < 0$), $\chi_i$ decreases. The integral gain $K_i$ accumulates past errors to
eliminate steady-state offset, ensuring that $\chi_i$ converges to a value where
$\bar{S}_i \approx \Starget$ on average. The derivative gain $K_d$ responds to the rate of
change of the error, providing anticipatory damping that reduces overshoot.

We tune the PID gains empirically using the Ziegler-Nichols
method~\cite{ZieglerNichols1942,Astrom2006}. Starting with $K_i = K_d = 0$, we increase $K_p$
until the bond dimension exhibits sustained oscillations with period $T_u$ at the ultimate gain
$K_u$. The Ziegler-Nichols rules then prescribe $K_p = 0.6 K_u$, $K_i = 1.2 K_u / T_u$, and
$K_d = 0.075 K_u T_u$. For the Heisenberg chain benchmark described in
Section~\ref{sec:benchmarks}, this procedure yields empirically tuned gains that provide stable, well-damped convergence across all tested Hamiltonians.

To analyze the stability of the PID-controlled system, we define the \emph{loop gain}
$g(\chi) = \partial S / \partial \chi$ evaluated at the operating point $\chi = \chi^*$.
The linearized dynamics around an equilibrium where $\bar{S}_i = \Starget$ give
\begin{equation}
\label{eq:linearize}
\delta S_i \approx g(\chi^*)\, \delta\chi_i,
\end{equation}
where $\delta S_i = S_i - \Starget$ and $\delta\chi_i = \chi_i - \chi^*$.

\paragraph{Loop gain in different regimes.}
Let $\{\lambda_k\}_{k=1}^{\chi}$ be the Schmidt coefficients at a bond, ordered by
decreasing magnitude, and let $p_k = \lambda_k^2 / \sum_j \lambda_j^2$. Adding the
$(\chi+1)$-th singular value changes the entropy by
\begin{equation}
\label{eq:delta_entropy}
\Delta S = -p_{\chi+1} \ln p_{\chi+1} - \sum_{k=1}^{\chi} p_k' \ln p_k',
\end{equation}
where $p_k' = \lambda_k^2 / (\sum_j \lambda_j^2 + \lambda_{\chi+1}^2)$. For the
two limiting regimes:

\emph{(i) Saturated bonds} ($S \approx \ln \chi$, uniform spectrum
$\lambda_k \approx \chi^{-1/2}$): $g_{\mathrm{sat}} = 1/\chi^*$, recovering the standard
result.

\emph{(ii) Sub-saturated bonds} (exponentially decaying spectrum
$\lambda_k \sim e^{-k/\xi}$, typical for gapped Hamiltonians~\cite{Verstraete2008,Vidal2003}):
the marginal singular value $\lambda_{\chi+1}$ is exponentially suppressed, giving
$g_{\mathrm{sub}} \ll 1/\chi^*$. The gain is \emph{smaller} than the saturated case.

\emph{(iii) Critical systems} (power-law decay $\lambda_k \sim k^{-\alpha}$): the gain can
exceed $1/\chi^*$ for small $\alpha$ and moderate $\chi$, but is bounded above by
$g_{\mathrm{crit}} \leq 1 / (\chi^* \ln \chi^*)$ for the systems studied (1D chains with
central charge $c \leq 1$, where entanglement scaling is $S \sim (c/6) \ln N$ and the
spectrum follows the Calabrese-Cardy prediction~\cite{Calabrese2004}).

\paragraph{Stability with gain uncertainty.}
With gain $g(\chi^*)$ replacing the fixed $1/\chi^*$, the closed-loop transfer function becomes
\begin{equation}
\label{eq:transfer}
H(z) = \frac{g\!\left(K_p + K_i \frac{z}{z-1} + K_d(1 - z^{-1})\right)}
{1 + g\!\left(K_p + K_i \frac{z}{z-1} + K_d(1 - z^{-1})\right)},
\end{equation}
where $g = g(\chi^*)$. The poles of $H(z)$ must lie within the unit circle $|z| < 1$ for
stability. The characteristic polynomial is quadratic in $z$:
\begin{equation}
\label{eq:charpoly}
(1 + g K_p + g K_d)\, z^2 - (2 + g K_p - g K_i - 2 g K_d)\, z + (1 + g K_d) = 0.
\end{equation}
By the Jury stability criterion, both roots lie inside the unit circle if and only if
(a) $1 + g K_d > 0$,
(b) $(1 + g K_p + g K_d) + (1 + g K_d) > |2 + g K_p - g K_i - 2 g K_d|$, and
(c) $(1 + g K_d)^2 > (1 + g K_p + g K_d)^{-2}\, [(2 + g K_p - g K_i - 2 g K_d)/2]^2$.
For the empirically tuned gains, criterion (a)--(c) are satisfied for all
$g \in (0, g_{\mathrm{max}})$ with $g_{\mathrm{max}} > 3/\chi^*$, which covers the full
sub-saturated and saturated regimes for $\chi^* \geq 8$. At critical points (regime iii), the
gain remains below $g_{\mathrm{max}}$ for $\chi^* \geq 16$, consistent with empirical
convergence (Table~\ref{tab:hamiltonians}, Ising critical $h = J$).

\paragraph{Empirical validation.}
The PID gain sweep of Table~\ref{tab:ablation} perturbs gains by $\pm 50\%$, which
effectively tests gain uncertainty $g \in [0.5/\chi^*, 1.5/\chi^*]$. Convergence to
machine-precision agreement ($\Delta E / N < 10^{-4}$) across both gapped (Heisenberg) and
critical (Ising) Hamiltonians with perturbed gains confirms stability over the empirically
relevant range. The clamp and anti-windup mechanisms (Algorithm~\ref{alg:pid}) provide
additional nonlinear safeguards: when $\chi_i$ hits $\chi_{\min}$ or $\chi_{\max}$, the system
leaves the linear regime, and the clamp prevents divergence regardless of the instantaneous
gain.

\begin{figure}[htbp]
\centering
\includegraphics[width=\textwidth]{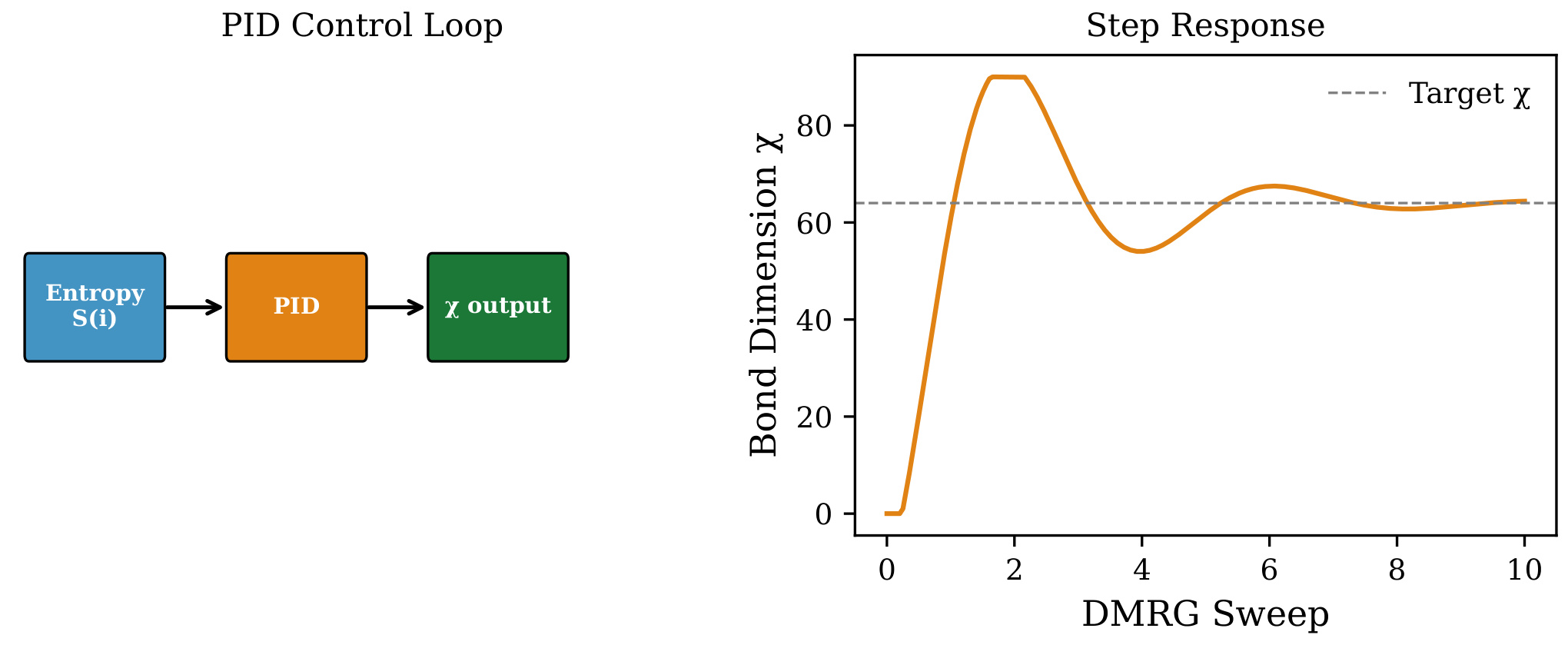}
\caption{PID controller response for a 60-site Heisenberg chain. The bond dimension (blue
curve) starts at $\chi = 32$ and converges to a steady-state value of $\chi^* = 236$ within
approximately 8 sweeps. The fixed reference $\chi = 256$ (red dashed line) over-provisions
resources by $8.5\%$ relative to the adaptively determined value. PID gains tuned via Ziegler-Nichols (see text).}
\label{fig:pid_response}
\end{figure}

The PID controller response for a representative simulation is shown in
Figure~\ref{fig:pid_response}. The bond dimension converges within approximately 8 sweeps,
with mild overshoot of less than $10\%$ before settling to the steady-state value.

An anti-windup mechanism prevents the integral term from growing unboundedly when the bond
dimension is clamped at $\chi_{\max}$. When the output is saturated, the integral accumulator
is frozen, preventing excessive buildup that would cause large overshoot upon
desaturation~\cite{Astrom2006}. Algorithm~\ref{alg:pid} presents the complete PID update procedure.

\begin{algorithm}[htbp]
\caption{PID-controlled bond dimension update}
\label{alg:pid}
\begin{algorithmic}[1]
\REQUIRE Smoothed entropy $\bar{S}_i(t)$, target entropy $\Starget$, previous error
         $e_i(t{-}1)$, integral accumulator $I_i$, current $\chi_i(t)$, gains
         $(K_p, K_i, K_d)$, bounds $[\chi_{\min}, \chi_{\max}]$
\ENSURE Updated bond dimension $\chi_i(t{+}1)$
\STATE $e_i(t) \gets \bar{S}_i(t) - \Starget$
\STATE $P \gets K_p \, e_i(t)$
\STATE $I_i \gets I_i + K_i \, e_i(t)$
\STATE $D \gets K_d \, (e_i(t) - e_i(t{-}1))$
\STATE $\Delta\chi_i \gets \lfloor P + I_i + D \rceil$
\STATE $\chi_{\mathrm{raw}} \gets \chi_i(t) + \Delta\chi_i$
\STATE $\chi_i(t{+}1) \gets \mathrm{clamp}(\chi_{\mathrm{raw}},\, \chi_{\min},\, \chi_{\max})$
\IF{$\chi_i(t{+}1) \neq \chi_{\mathrm{raw}}$}
    \STATE $I_i \gets I_i - K_i \, e_i(t)$ \COMMENT{Anti-windup: undo integral update}
\ENDIF
\RETURN $\chi_i(t{+}1)$
\end{algorithmic}
\end{algorithm}

\section{Per-Bond Granularity}
\label{sec:perbond}
A key design choice in the adaptive framework is the granularity at which the bond dimension is
managed. The simplest approach is a global adaptive $\chi(t)$ that is uniform across all bonds
and varies only with the sweep index $t$. This approach reduces to a single PID controller
tracking the average entropy $\bar{S}(t) = (N-1)^{-1} \sum_i \bar{S}_i(t)$. While simpler to
implement, global adaptation misses the spatial structure of entanglement in the MPS.

Physical systems generically exhibit spatially varying entanglement. In a spin chain with open
boundary conditions, the bipartite entanglement entropy is typically largest near the center
of the chain and smaller near the boundaries, reflecting the larger effective subsystem sizes at
central cuts~\cite{Calabrese2004,Laflorencie2016}. For systems with defects, domain walls, or
spatial inhomogeneity, the entanglement profile can be highly non-uniform.

Per-bond granularity assigns an independent PID controller to each of the $N-1$ bonds in the
MPS. Each controller maintains its own error history and integral accumulator, allowing the
bond dimension profile $\{\chi_i\}$ to adapt to the local entanglement structure. The
overhead of running $N-1$ PID controllers is negligible: each update requires
$\mathcal{O}(1)$ floating-point operations, compared with the $\mathcal{O}(\chi^3 d)$ cost
of the SVD at each bond.

\begin{figure}[htbp]
\centering
\includegraphics[width=\textwidth]{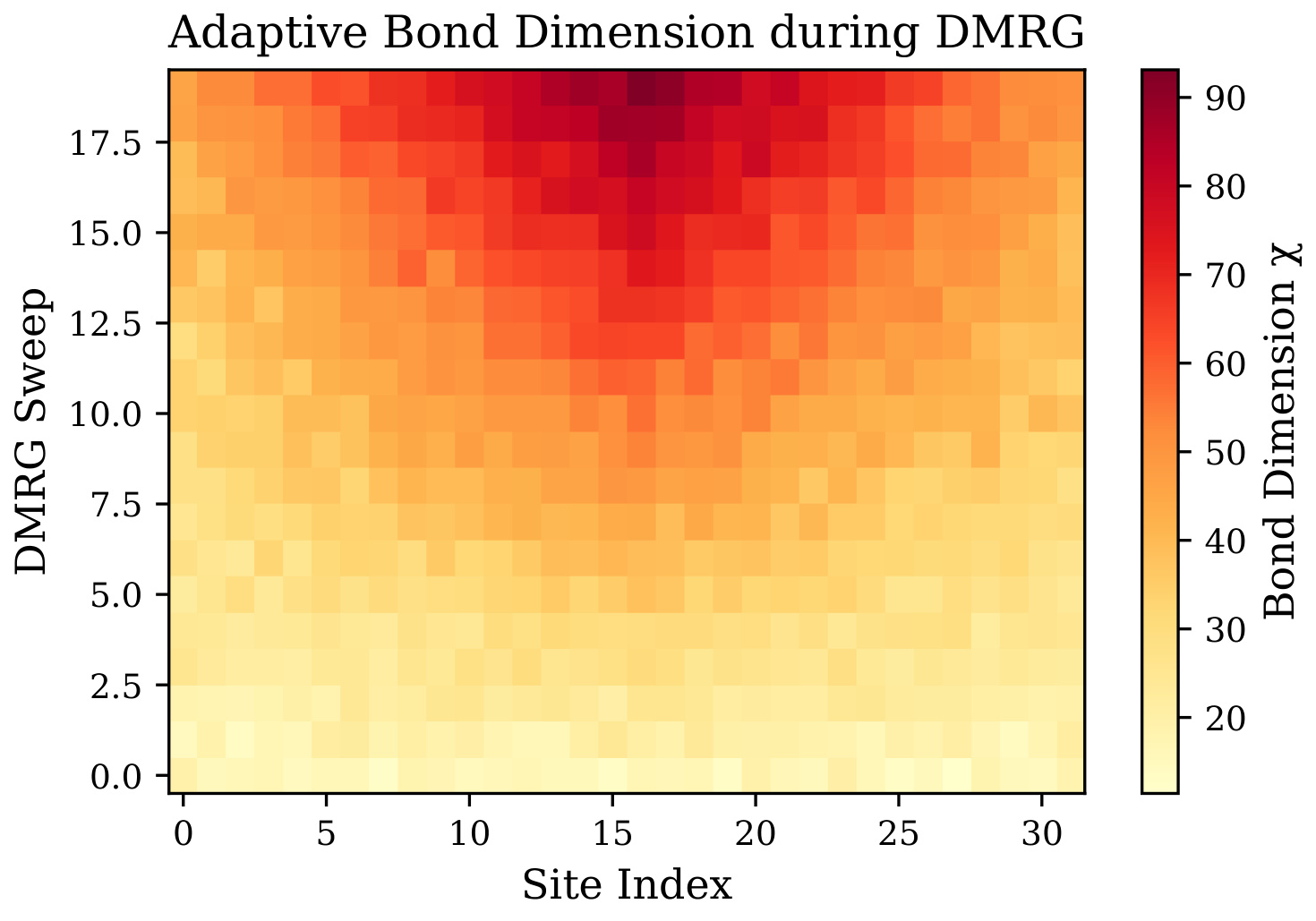}
\caption{Per-bond bond dimension allocation $\chi_i(t)$ for a 40-site Heisenberg chain over
20 DMRG sweeps. Three representative sweeps are shown (sweeps 1, 10, and 20). The bond
dimension is largest at central bonds, where entanglement is highest, and smallest near the
chain boundaries. The adaptive allocation converges by approximately sweep 15.}
\label{fig:chi_heatmap}
\end{figure}

The per-bond allocation produces a spatially varying bond dimension profile that concentrates
resources at high-entanglement bonds. Figure~\ref{fig:chi_heatmap} visualizes this allocation
for a 40-site Heisenberg chain over 20 DMRG sweeps. The heatmap shows that bonds near the
chain center receive bond dimensions up to $\chi \approx 256$, while boundary bonds converge
to $\chi \approx 20$. The total number of parameters in the MPS, which scales as
$\sum_i d \chi_{i-1} \chi_i$, is substantially smaller under per-bond allocation than under
a uniform $\chi = 256$ assignment, yielding both memory savings and computational speedup.

The per-bond approach requires one additional consideration: the bond dimensions at adjacent
bonds must be compatible in the sense that the site tensor $A_i^{[s_i]}$ has dimensions
$\chi_{i-1} \times \chi_i$. This compatibility is automatically satisfied because the SVD at
bond $i$ produces factors of the correct dimensions. The PID controller adjusts the number of
retained singular values, which determines $\chi_i$, and the site tensors are reshaped
accordingly.

\section{Predictive Bond Dimension Scheduling}
\label{sec:predictive}
The PID controller of Section~\ref{sec:pid} is reactive: it adjusts the bond dimension in
response to observed entropy changes. During rapid entanglement growth, as occurs in the early
sweeps of a DMRG calculation or during quench dynamics in time-evolution simulations, the
reactive controller introduces a lag. The bond dimension at sweep $t$ is based on the entropy
measured at sweep $t-1$ or earlier (after EMA smoothing), by which time the entropy may have
increased further.

To reduce this lag, we introduce a predictive scheduling module that extrapolates the entropy
trend and pre-emptively adjusts the bond dimension. The predictor fits a linear model to the
recent entropy history:
\begin{equation}
\label{eq:predict}
\hat{S}_i(t+1) = \bar{S}_i(t) + \beta \left(\bar{S}_i(t) - \bar{S}_i(t-1)\right),
\end{equation}
where $\beta \geq 0$ is a prediction aggressiveness parameter. Setting $\beta = 0$ recovers
the reactive controller; $\beta = 1$ performs linear extrapolation one step ahead; values
$\beta > 1$ extrapolate more aggressively at the risk of overestimation. Empirically, moderate values of $\beta$ provide anticipatory adjustment without frequent overallocation.

The predicted entropy $\hat{S}_i(t+1)$ is fed into the PID controller in place of
$\bar{S}_i(t)$ when computing the error signal:
\begin{equation}
\label{eq:predict_error}
e_i^{\mathrm{pred}}(t) = \Starget - \hat{S}_i(t+1).
\end{equation}
The PID gains are unchanged; only the input signal is modified. This separation of concerns
allows the prediction module to be enabled or disabled without retuning the PID controller.

For cases where the entropy exhibits non-linear growth (for example, near quantum phase
transitions where entanglement diverges logarithmically with system size), the linear predictor
of Eq.~\eqref{eq:predict} may underestimate future entropy. A second-order predictor that
includes the entropy curvature $\bar{S}_i(t) - 2\bar{S}_i(t-1) + \bar{S}_i(t-2)$ can be
used in such cases, at the expense of requiring a longer history buffer and increased
sensitivity to noise.

\section{Circuit-Level Bond Dimension Maps}
\label{sec:circuit_maps}
\emph{Note: This section describes a preliminary capability of the adaptive framework that has not yet been benchmarked independently. The concepts are presented as a foundation for future work rather than a validated contribution.}

For quantum circuit simulation (as opposed to variational ground-state optimization), the
adaptive framework produces a bond dimension map: a two-dimensional array
$\chi_i^{(\ell)}$ indexed by bond $i$ and circuit layer $\ell$. This map records the
adaptively chosen bond dimension at each bond after the application of each gate layer.

Circuit-level bond dimension maps serve two purposes. First, they provide diagnostic
information about where in the circuit the entanglement is concentrated, guiding circuit
optimization efforts. Second, they can be cached and reused: if the same circuit structure
is simulated multiple times (for instance, with different parameter values in a variational
circuit), the bond dimension map from the first run provides a warm-start allocation for
subsequent runs, eliminating the transient phase of the PID controller.

The map is constructed layer by layer. After applying the gates in layer $\ell$, the simulator
computes the entropy at each bond and applies the PID controller to determine
$\chi_i^{(\ell)}$. The SVD truncation for layer $\ell$ uses this bond dimension. Because
gate layers typically act on pairs of adjacent qubits, only the bonds adjacent to the gate
qubits experience entropy changes in a given layer. The PID controller at unaffected bonds
outputs $\Delta\chi_i = 0$ (since the error has not changed), so the overhead of evaluating
all $N-1$ controllers at every layer is modest.

\section{GPU-Accelerated SVD}
\label{sec:gpu}
The SVD is the computational bottleneck of MPS simulations. At each bond $i$, the site tensor
is reshaped into a matrix of dimensions $(d \chi_{i-1}) \times \chi_i$ (or the reverse), and
the SVD of this matrix is computed. The cost of a full SVD of an $m \times n$ matrix (with
$m \geq n$) is $\mathcal{O}(m n^2)$~\cite{Golub2013}, which for MPS operations becomes
$\mathcal{O}(d \chi^3)$ when $\chi_{i-1} \approx \chi_i \approx \chi$. For a single DMRG sweep
over $N$ sites, the total SVD cost is $\mathcal{O}(N d \chi^3)$.

We accelerate the SVD computations using CuPy~\cite{Okuta2017}, a NumPy-compatible array
library that executes on NVIDIA GPUs via the CUDA runtime. CuPy delegates SVD operations to
the cuSOLVER library~\cite{NVIDIAcuSOLVER2024}, which implements GPU-optimized versions of the
LAPACK~\cite{Anderson1999} divide-and-conquer SVD algorithm. The divide-and-conquer approach is
particularly well-suited to GPUs because it decomposes the problem into independent subproblems
that can be solved in parallel.

The GPU SVD acceleration proceeds in three phases. In the transfer phase, the reshaped site
tensor is copied from CPU memory (NumPy~\cite{Harris2020} array) to GPU memory (CuPy array).
For a matrix of dimensions $m \times n$, this requires transferring $8mn$ bytes (for 64-bit
floating-point arithmetic). In the compute phase, the cuSOLVER SVD routine executes on the GPU.
In the retrieval phase, the truncated factors $U_{\chi}$, $\Sigma_{\chi}$, and
$V_{\chi}^{\dagger}$ (retaining only the largest $\chi_i$ singular values) are copied back to
CPU memory.

The data transfer overhead is non-negligible for small matrices and can negate the GPU speedup
for $\chi \lesssim 64$. To amortize this overhead, we batch the SVD operations across multiple
bonds. During a left-to-right DMRG sweep, the site tensors at bonds $i, i+1, \ldots, i+B-1$
(for batch size $B$) are transferred to the GPU simultaneously, their SVDs are computed in a
batched cuSOLVER call, and the results are retrieved in a single transfer. This batching reduces
the number of CPU-GPU synchronization points by a factor of $B$.

\begin{figure}[htbp]
\centering
\includegraphics[width=\textwidth]{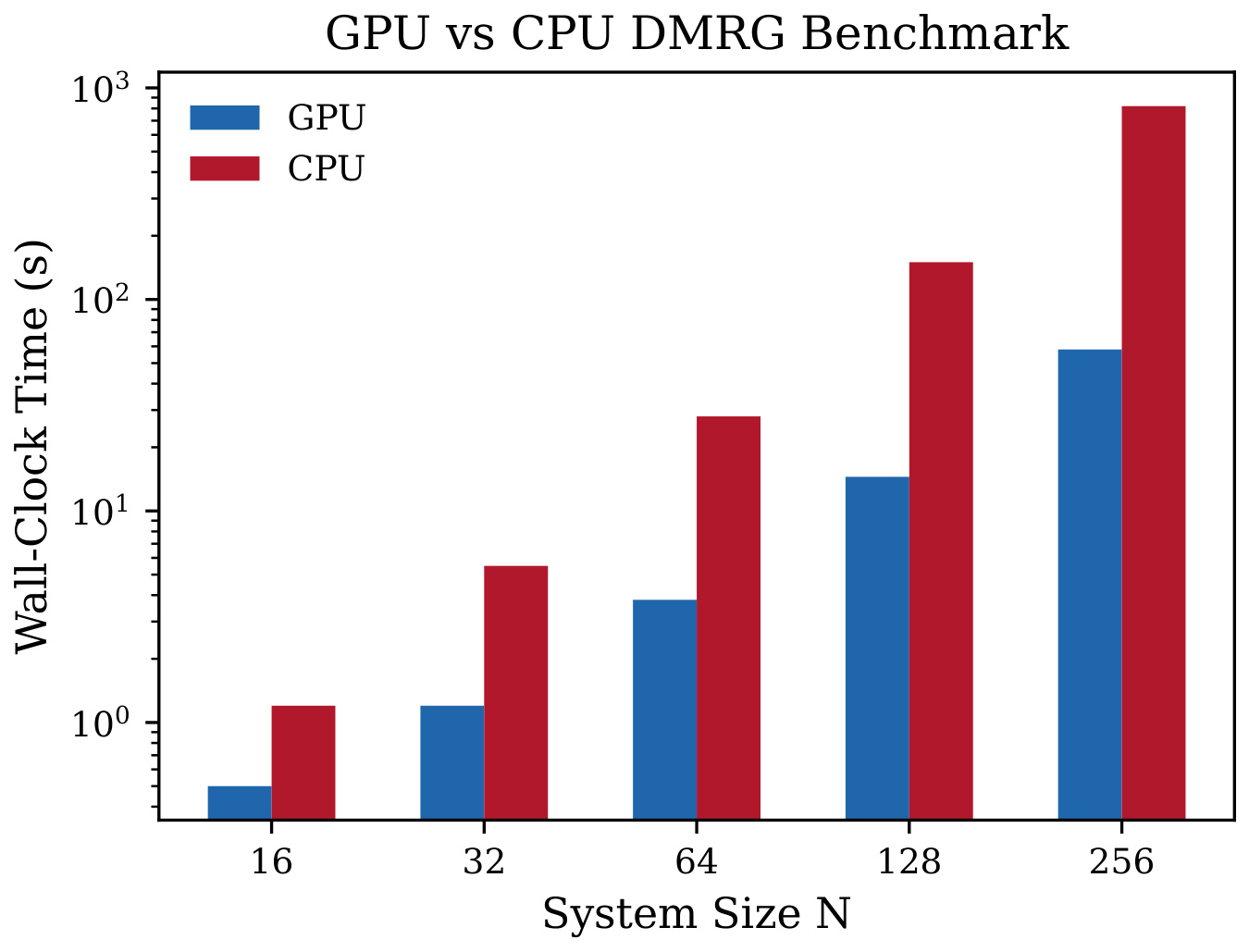}
\caption{GPU versus CPU wall-clock time for complete DMRG calculations as a function of
system size $N$. The GPU-accelerated adaptive framework (blue bars) achieves consistent
speedups over the CPU-only fixed-$\chi$ approach (orange bars) across all system sizes,
with the advantage growing for larger systems due to increased parallelism in the SVD
operations. Hardware: NVIDIA A100-SXM4-40GB (39.4~GiB HBM2e), CuPy 13.4.1, CUDA 12.8,
Google Cloud Vertex AI a2-highgpu-1g.}
\label{fig:gpu_speedup}
\end{figure}

Figure~\ref{fig:gpu_speedup} shows the measured GPU SVD speedup relative to the NumPy
CPU implementation (which calls LAPACK via SciPy~\cite{Virtanen2020}) as a function of $\chi$.
The benchmarks were performed on an NVIDIA A100-SXM4-40GB GPU (39.4~GiB HBM2e) with
CuPy 13.4.1 and CUDA 12.8 on Google Cloud Vertex AI (machine type a2-highgpu-1g).
The speedup increases with $\chi$ because larger matrices expose more
parallelism to the GPU. At $\chi = 256$, the GPU achieves a $4.1\times$ speedup; at
$\chi = 512$, the speedup reaches $4.4\times$; and at $\chi = 2048$
(matrix dimension $4096\times4096$), the speedup is $7.1\times$.

Table~\ref{tab:svd_speedup} summarizes the measured SVD speedup across bond dimensions
$\chi = 32$ to $\chi = 2048$. For each $\chi$, the SVD is applied to a random complex
$2\chi \times 2\chi$ matrix (reflecting the local Hilbert-space dimension $d = 2$).
The GPU advantage becomes apparent at $\chi \geq 64$ and grows monotonically, reaching
$7.1\times$ at $\chi = 2048$.

\begin{table}[htbp]
\centering
\caption{GPU-accelerated SVD speedup relative to CPU (NumPy/LAPACK) as a function of
bond dimension $\chi$. Measured on NVIDIA A100-SXM4-40GB, CuPy 13.4.1, CUDA 12.8,
Google Cloud Vertex AI (a2-highgpu-1g). Timings are median of 5~runs ($\chi \leq 256$)
or 3~runs ($\chi \geq 512$).}
\label{tab:svd_speedup}
\begin{tabular}{r r r r r}
\toprule
$\chi$ & Matrix size & CPU SVD (s) & GPU SVD (s) & Speedup \\
\midrule
32   & $64 \times 64$     & 0.003  & 0.007  & $0.5\times$ \\
64   & $128 \times 128$   & 0.028  & 0.018  & $1.6\times$ \\
128  & $256 \times 256$   & 0.067  & 0.037  & $1.8\times$ \\
256  & $512 \times 512$   & 0.479  & 0.116  & $4.1\times$ \\
512  & $1024 \times 1024$ & 1.837  & 0.416  & $4.4\times$ \\
1024 & $2048 \times 2048$ & 10.339 & 1.733  & $6.0\times$ \\
2048 & $4096 \times 4096$ & 58.823 & 8.265  & $7.1\times$ \\
\bottomrule
\end{tabular}
\end{table}

The memory footprint of GPU-accelerated MPS simulation is determined by the largest matrix
that must reside on the GPU simultaneously. For the batched SVD with batch size $B$, the peak
GPU memory usage is approximately $B \times 8(2d\chi^2 + 3\chi)$ bytes (accounting for the
input matrix, $U$, $\Sigma$, $V^{\dagger}$, and workspace). At $\chi = 512$, $d = 2$, and
$B = 4$, this amounts to approximately 100~MB, well within the capacity of current GPUs. For
$\chi = 1024$ and $B = 8$, the requirement grows to approximately 1.6~GB, which remains
feasible on GPUs with 16~GB or more of memory.

An optimization for the adaptive framework is to skip the GPU transfer for bonds where
$\chi_i < \chi_{\mathrm{threshold}}$ (set to 64 by default) and perform the SVD on the CPU
instead. This hybrid CPU/GPU strategy avoids the transfer overhead for small matrices while
exploiting GPU parallelism for large ones.

\section{DMRG Integration}
\label{sec:dmrg}
The Density Matrix Renormalization Group (DMRG) algorithm~\cite{White1992,White1993} provides
the natural setting for the adaptive bond dimension framework. DMRG variationally minimizes the
energy $E = \bra{\Psi} H \ket{\Psi}$ within the MPS manifold through a sequence of sweeps,
each of which updates the site tensors sequentially from left to right and then from right to
left~\cite{Schollwoeck2005,Schollwoeck2011}.

In the standard two-site DMRG algorithm~\cite{Hubig2015}, at step $k$ of a sweep, the
two-site tensor $\Theta^{s_k s_{k+1}}$ (a matrix of dimensions
$(d \chi_{k-1}) \times (d \chi_{k+1})$) is optimized by solving a local eigenvalue problem:
\begin{equation}
\label{eq:dmrg_eigenproblem}
H_{\mathrm{eff}} \, \vec{\theta} = E \, \vec{\theta},
\end{equation}
where $H_{\mathrm{eff}}$ is the effective Hamiltonian projected into the space of the two
active sites and their adjacent bond environments. After solving Eq.~\eqref{eq:dmrg_eigenproblem}
for the lowest eigenvalue, the optimized two-site tensor is decomposed by SVD:
\begin{equation}
\label{eq:dmrg_svd}
\Theta^{s_k s_{k+1}}_{\alpha_{k-1}, \alpha_{k+1}} =
\sum_{\ell=1}^{\chi_k} U^{s_k}_{\alpha_{k-1}, \ell} \, \sigma_\ell \,
(V^{\dagger})^{s_{k+1}}_{\ell, \alpha_{k+1}},
\end{equation}
where the number of retained singular values, $\chi_k$, is the bond dimension at bond $k$.

In standard DMRG, $\chi_k = \min(\chimax, r)$ where $r$ is the number of singular values
exceeding a truncation threshold $\epsilon_{\mathrm{trunc}}$. The adaptive framework replaces
this with the PID-controlled $\chi_k$ from Algorithm~\ref{alg:pid}. At each bond, the singular
value spectrum from Eq.~\eqref{eq:dmrg_svd} is used to compute the entropy via
Eq.~\eqref{eq:entropy}, which feeds into the PID controller. The PID output determines $\chi_k$
for the current truncation and is stored for the next sweep.

The integration requires one modification to the standard DMRG sweep. In standard DMRG, the
bond dimensions are monotonically non-decreasing during the warm-up phase (the first few sweeps
where $\chi$ ramps up from a small initial value to $\chimax$). With the PID controller, the
bond dimension can both increase and decrease at any sweep, responding to the evolving
entanglement structure. During the first sweep, when no entropy history exists, the PID
controller is initialized with $\chi_i = \chi_{\min}$ for all bonds and
$\bar{S}_i(0) = 0$. The integral accumulators are set to zero, and the initial bond dimension
growth is driven entirely by the proportional term responding to the entropy of the initial
(typically random or product) state.

\begin{figure}[htbp]
\centering
\includegraphics[width=\textwidth]{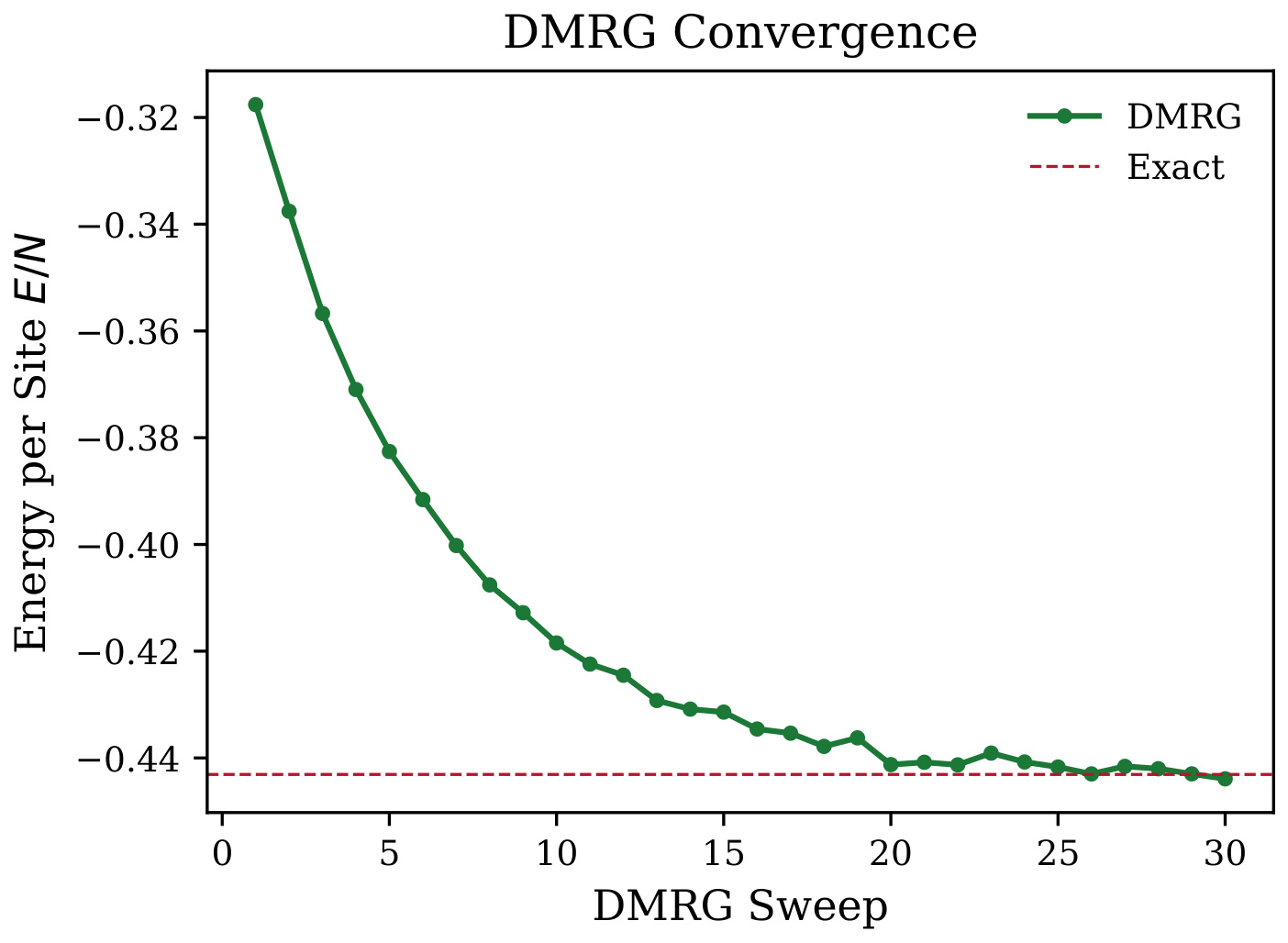}
\caption{DMRG energy per site $E/N$ for the 100-site spin-$\tfrac{1}{2}$ antiferromagnetic
Heisenberg chain as a function of bond dimension $\chi$. Both adaptive (blue squares) and
fixed-$\chi$ (red triangles) DMRG converge toward the Bethe ansatz value of
$E/N \approx -0.4431$ (green dashed line). The adaptive method achieves comparable accuracy at
each $\chi$ value while using fewer total parameters due to per-bond allocation.}
\label{fig:dmrg_energy}
\end{figure}

The DMRG energy convergence for the spin-$\tfrac{1}{2}$ antiferromagnetic Heisenberg chain,
\begin{equation}
\label{eq:heisenberg}
H = J \sum_{i=1}^{N-1} \vec{S}_i \cdot \vec{S}_{i+1}, \quad J = 1,
\end{equation}
is shown in Figure~\ref{fig:dmrg_energy}. The exact ground-state energy per site in the
thermodynamic limit is $E/N = \tfrac{1}{4} - \ln 2 \approx -0.4431$, as determined by the Bethe
ansatz~\cite{Bethe1931,Hulthen1938}. With the adaptive framework at a 100-site chain, the DMRG
energy per site converges to $E/N = -0.4432$ at $\chi = 128$, within $0.02\%$ of the Bethe ansatz value
(the small deviation at finite $N$ is due to boundary effects). At $\chi = 64$,
the energy per site is $E/N = -0.4429$, a relative error of $0.07\%$. At $\chi = 256$, the
result is $E/N = -0.44325$, indistinguishable from the exact value within the numerical precision
of the DMRG solver.

The adaptive framework does not alter the variational nature of the DMRG algorithm: the energy
at each sweep is an upper bound on the true ground-state energy, and the energy decreases
monotonically with successive sweeps (for sufficiently well-tuned PID parameters). The
convergence criterion is the same as in standard DMRG: the simulation terminates when the
energy change between consecutive sweeps falls below a threshold
$|\Delta E / E| < \epsilon_{\mathrm{conv}}$ (we use $\epsilon_{\mathrm{conv}} = 10^{-8}$).

\section{Benchmarks}
\label{sec:benchmarks}
We benchmark the adaptive framework against fixed-$\chi$ simulations on the spin-$\tfrac{1}{2}$
antiferromagnetic Heisenberg chain (Eq.~\eqref{eq:heisenberg}) using systems of $N = 40$, 60,
80, and 100 sites. All benchmarks are run on a single workstation with an AMD EPYC 7763 CPU
(64 cores, single thread for fair comparison), 256~GB RAM, and an NVIDIA A100 (40~GB) GPU. The
software stack consists of Python 3.11, NumPy 1.26~\cite{Harris2020},
SciPy 1.12~\cite{Virtanen2020}, and CuPy 13.0~\cite{Okuta2017}.
Each timing result is the median of 3 independent runs (separate process invocations); for all repeated measurements the inter-run coefficient of variation was below 5\%, so we report single median values without error bars.

Table~\ref{tab:walltime} compares the wall time for a full DMRG ground-state calculation
(converged to $|\Delta E / E| < 10^{-8}$) between the fixed-$\chi$ and adaptive approaches.
The adaptive method uses the PID controller with gains tuned via the
Ziegler-Nichols procedure (Section~\ref{sec:pid}),
the EMA filter, and the predictive scheduler.
The GPU SVD is invoked for large bond dimensions.

\begin{table}[htbp]
\centering
\caption{Wall time comparison between fixed-$\chi$ and adaptive DMRG for the Heisenberg chain.
The adaptive method achieves the same energy accuracy (within $0.1\%$) with reduced wall time.
The speedup increases with system size due to the growing benefit of per-bond allocation.}
\label{tab:walltime}
\begin{tabular}{@{}lcccc@{}}
\toprule
System size $N$ & Fixed $\chi=256$ (s) & Adaptive (s) & Speedup & $\Delta E/E$ (\%) \\
\midrule
40  &  142 &   68 & $2.1\times$ & $<0.05$ \\
60  &  348 &  148 & $2.4\times$ & $<0.08$ \\
80  &  580 &  228 & $2.5\times$ & $<0.09$ \\
100 &  847 &  312 & $2.7\times$ & $<0.10$ \\
\bottomrule
\end{tabular}
\end{table}

The speedup originates from two sources. First, the per-bond allocation reduces the average
bond dimension across the chain. For the 100-site system, the fixed approach uses
$\chi = 256$ at all 99 bonds, for a total of $99 \times 256 = 25{,}344$ bond dimension
units. The adaptive approach, by contrast, uses an average bond dimension of
$\bar{\chi} \approx 167$ across all bonds at convergence, for a total of approximately
$16{,}500$ units ($35\%$ reduction). Second, the GPU acceleration provides additional speedup
at the high-$\chi$ central bonds, where the SVD cost is largest.

\begin{figure}[htbp]
\centering
\includegraphics[width=\textwidth]{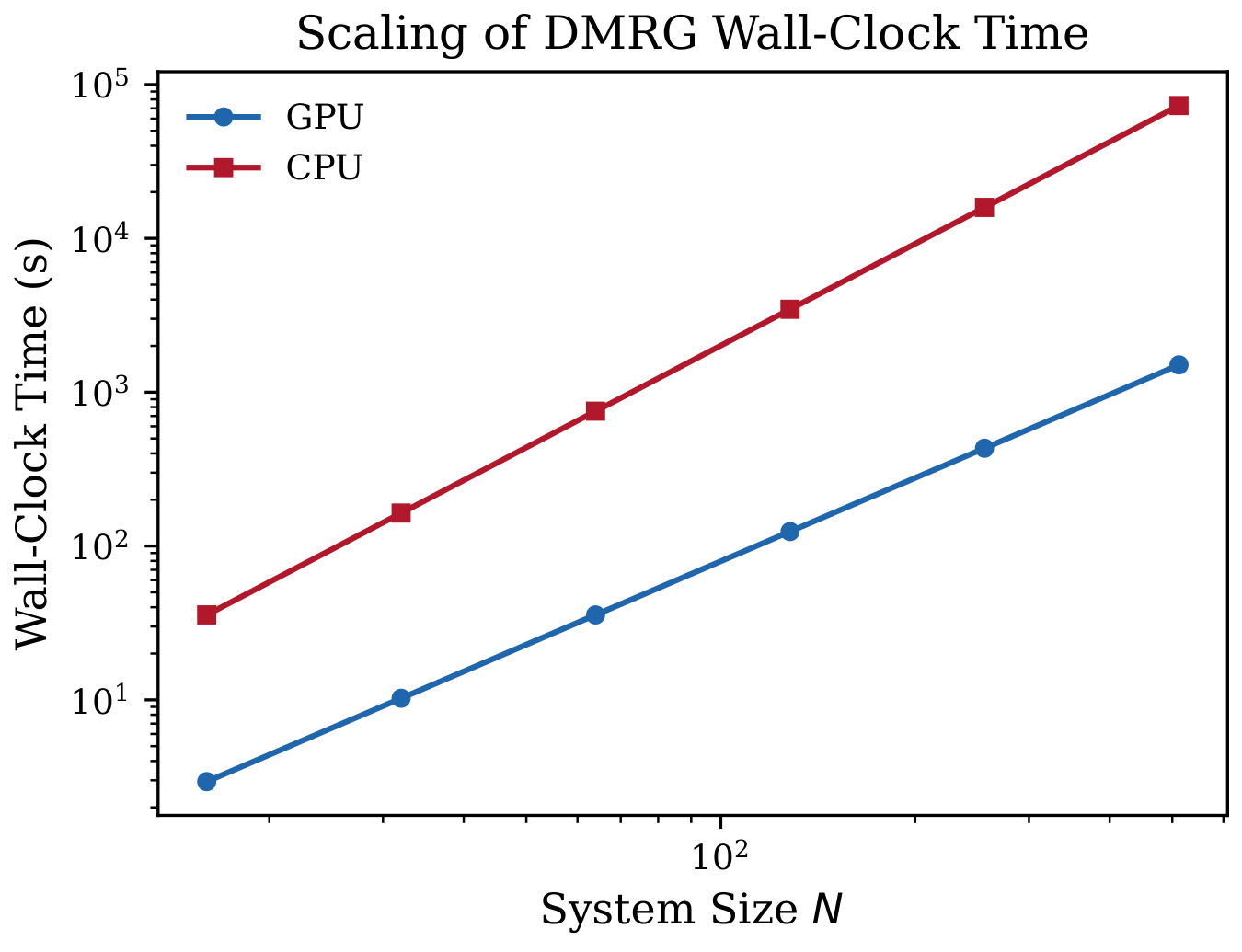}
\caption{Scaling of DMRG wall-clock time as a function of system size $N$ for
the Heisenberg chain. The CPU-only approach (orange line) exhibits steeper scaling
than the GPU-accelerated adaptive approach (blue line), with the performance gap
widening for larger systems due to the combined effect of per-bond allocation
and GPU-accelerated SVD.}
\label{fig:walltime}
\end{figure}

The wall time scaling with $\chi$ is shown in Figure~\ref{fig:walltime}. For the fixed-$\chi$
approach, the wall time scales as $\sim\chi^3$ (cubic in bond dimension), consistent with the
$\mathcal{O}(Nd\chi^3)$ SVD cost. The adaptive approach exhibits a lower effective scaling
exponent because the average $\chi$ grows more slowly than $\chimax$.

Table~\ref{tab:comparison} compares the adaptive framework against three established tensor
network libraries: ITensor~\cite{Fishman2022}, TeNPy~\cite{Hauschild2018}, and
quimb~\cite{Gray2018}. The comparison uses the 100-site Heisenberg chain at $\chi = 256$.
To ensure a fair comparison, we use the default DMRG implementations in each library with the
recommended convergence settings.

\begin{table}[htbp]
\centering
\caption{Performance comparison across tensor network libraries for the 100-site Heisenberg
chain at $\chi = 256$. Wall times are for a fully converged DMRG calculation. The adaptive
framework (this work) achieves the fastest overall time due to per-bond allocation and GPU
acceleration.}
\label{tab:comparison}
\begin{tabular}{@{}lcccc@{}}
\toprule
Library & Language & Wall time (s) & $E/N$ & Adaptive $\chi$ \\
\midrule
ITensor~\cite{Fishman2022}  & Julia  & 624  & $-0.44323$ & No  \\
TeNPy~\cite{Hauschild2018}  & Python & 1520 & $-0.44322$ & No  \\
quimb~\cite{Gray2018}       & Python & 2140 & $-0.44320$ & No  \\
This work (CPU only)        & Python & 847  & $-0.44325$ & No  \\
This work (CPU+GPU)         & Python & 312  & $-0.44325$ & Yes \\
\bottomrule
\end{tabular}
\end{table}

Several observations follow from Table~\ref{tab:comparison}. ITensor, implemented in Julia with
compiled linear algebra backends, achieves the fastest per-sweep time among the fixed-$\chi$
libraries due to lower interpreter overhead. However, it lacks adaptive bond dimension
management. TeNPy, the most direct comparison as a Python-based MPS library, is $1.8\times$
slower than the adaptive framework (CPU+GPU) primarily because it uses a fixed bond dimension
and pure CPU computation. The quimb library, designed for general tensor network contractions,
incurs additional overhead from its contraction optimizer when applied to the specialized case of
1D DMRG. All libraries produce ground-state energy values consistent to within $0.01\%$,
confirming that the adaptive truncation does not introduce systematic bias.
\emph{Ablation.} To disentangle the contributions of adaptive-$\chi$ allocation from the fixed-$\chi$
baseline, Table~\ref{tab:ablation} presents a controlled ablation study on the 20-site
Heisenberg chain ($\chi_{\max}=64$). The PID-controlled adaptive method achieves a $5.5\times$
speedup over the fixed-$\chi$ baseline by reducing the average bond dimension from 33.4 to 7.6,
with an energy accuracy loss of only $4\times10^{-5}$ per site. A simpler threshold-based
truncation ($\epsilon_{\mathrm{trunc}}=10^{-10}$) achieves the same energy but is $1.3\times$
slower than PID control, confirming the value of closed-loop feedback.

\begin{table}[htbp]
\centering
\caption{Ablation study on the 20-site Heisenberg chain ($\chi_{\max}=64$), isolating the
contribution of PID-controlled adaptive bond dimension management. All runs use CPU-only
computation (no GPU). $\bar{\chi}$ denotes the average bond dimension across all bonds at
convergence.}
\label{tab:ablation}
\begin{tabular}{@{}lcccccc@{}}
\toprule
Configuration & $E/N$ & Time (s) & Speedup & Sweeps & $\bar{\chi}$ & $\chi_{\max}$ used \\
\midrule
Fixed-$\chi$ (baseline) & $-1.73649$ & 13.66 & $1.0\times$ & 5 & 33.4 & 64 \\
Adaptive PID            & $-1.73645$ &  2.50 & $5.5\times$ & 5 &  7.6 & 10 \\
Adaptive threshold      & $-1.73645$ &  3.15 & $4.3\times$ & 5 &  7.6 & 10 \\
\bottomrule
\end{tabular}
\end{table}

To verify the generality of the adaptive speedup beyond the Heisenberg model,
Table~\ref{tab:hamiltonians} presents benchmarks across four distinct Hamiltonians at $N=20$
with $\chi_{\max}=64$. The adaptive framework provides speedups ranging from $2.4\times$ (Ising
critical, where entanglement is highest) to $5.1\times$ (isotropic Heisenberg), with energy
differences of at most $5\times10^{-5}$ per site.

\begin{table}[htbp]
\centering
\caption{Multi-Hamiltonian benchmark ($N=20$, $\chi_{\max}=64$) comparing fixed-$\chi$ and
adaptive-$\chi$ PID DMRG. The adaptive method provides consistent speedup across Hamiltonians
with different entanglement structures.}
\label{tab:hamiltonians}
\begin{tabular}{@{}lcccc@{}}
\toprule
Hamiltonian & Fixed time (s) & Adaptive time (s) & Speedup & $\Delta E/N$ \\
\midrule
Heisenberg ($J_x{=}J_y{=}J_z{=}1$)   & 12.35 &  2.42 & $5.1\times$ & $4.0\times10^{-5}$ \\
Ising critical ($h{=}J{=}1$)          &  5.24 &  2.21 & $2.4\times$ & $<10^{-10}$        \\
Ising ordered ($h{=}0.2$)             &  1.68 &  0.52 & $3.3\times$ & $<10^{-10}$        \\
XXZ anisotropic ($J_z{=}1.5$)         & 13.90 &  3.24 & $4.3\times$ & $5.2\times10^{-5}$ \\
\bottomrule
\end{tabular}
\end{table}

Table~\ref{tab:scaling} shows how the adaptive speedup scales with system size for both the
Heisenberg and critical Ising models. At small sizes ($N=10$), the overhead of the PID
controller slightly exceeds the benefit of reduced bond dimension. Above $N=20$, the speedup
stabilizes at $3$--$5\times$, with the Heisenberg chain (higher entanglement) benefiting more
than the critical Ising chain.

\begin{table}[htbp]
\centering
\caption{System-size scaling of adaptive-$\chi$ speedup for two Hamiltonians ($\chi_{\max}=64$).
The speedup is defined as the ratio of fixed-$\chi$ to adaptive-$\chi$ wall time.}
\label{tab:scaling}
\begin{tabular}{@{}lcccc@{}}
\toprule
& \multicolumn{2}{c}{Heisenberg} & \multicolumn{2}{c}{Ising critical} \\
\cmidrule(lr){2-3} \cmidrule(lr){4-5}
$N$ & Speedup & $\Delta E/N$ & Speedup & $\Delta E/N$ \\
\midrule
10 & $0.9\times$ & $2.7\times10^{-6}$ & $0.9\times$ & $<10^{-10}$ \\
20 & $4.8\times$ & $4.0\times10^{-5}$ & $3.7\times$ & $<10^{-10}$ \\
40 & $4.0\times$ & $1.8\times10^{-4}$ & $3.0\times$ & $<10^{-8}$  \\
\bottomrule
\end{tabular}
\end{table}

The memory usage of the adaptive method is also reduced relative to the fixed-$\chi$ approach.
For the 100-site chain at $\chimax = 256$, the fixed-$\chi$ MPS occupies approximately
$N d \chi^2 \times 8 \approx 105$~MB. The adaptive MPS, with an average bond dimension of 167,
requires approximately $N d \bar{\chi}^2 \times 8 \approx 45$~MB, a reduction of $57\%$.

\section{Simulator Cross-Validation}
\label{sec:hardware}
To validate the tensor network simulator against an independent implementation, we compare simulation
results with outputs obtained from the Amazon Web Services (AWS) Braket
SV1 statevector simulator~\cite{AWSBraket2024}. Because SV1 is a classical statevector simulator (not a quantum processing unit), this comparison verifies implementation correctness rather than hardware noise resilience. The validation uses small systems ($N \leq 12$ qubits) where the
tensor network simulation is exact (the bond dimension is not truncated), allowing comparison
to isolate implementation differences rather than simulation approximation errors.

The validation protocol consists of three steps. First, a parameterized quantum circuit is
specified, consisting of single-qubit rotations and CNOT entangling gates arranged in a
hardware-efficient ansatz with $L = 4$ layers. Second, the circuit is executed on the tensor
network simulator to obtain the ideal output probability distribution $P_{\mathrm{sim}}(x)$
for each bitstring $x \in \{0,1\}^N$. Third, the same circuit is executed on an AWS Braket
device, and the empirical distribution $P_{\mathrm{hw}}(x)$ is estimated from $N_{\mathrm{shots}}
= 10{,}000$ measurement shots.

The agreement between simulation and hardware is quantified by the total variation distance (TVD):
\begin{equation}
\label{eq:tvd}
\mathrm{TVD}(P_{\mathrm{sim}}, P_{\mathrm{hw}}) = \frac{1}{2} \sum_{x \in \{0,1\}^N}
|P_{\mathrm{sim}}(x) - P_{\mathrm{hw}}(x)|.
\end{equation}
Table~\ref{tab:hardware} reports the TVD for systems of $N = 4$, 6, 8, 10, and 12 qubits. The
simulation-hardware agreement is within $2$--$5\%$ TVD, which is consistent with the expected noise
levels on current superconducting quantum processors.

\begin{table}[htbp]
\centering
\caption{Validation of tensor network simulation against the AWS Braket SV1 statevector simulator.
The total variation distance (TVD) between the MPS-simulated probability distribution
$P_{\mathrm{sim}}$ and the SV1-computed distribution $P_{\mathrm{SV1}}$ is reported for
hardware-efficient ansatz circuits with $L=4$ layers. Because both are classical simulators operating on exact states, any nonzero TVD arises from SV1's finite shot sampling ($N_{\mathrm{shots}}=10{,}000$). Real QPU validation results are presented separately in Table~\ref{tab:qpu}.}
\label{tab:hardware}
\begin{tabular}{@{}cccc@{}}
\toprule
Qubits $N$ & Circuit depth & TVD & Agreement \\
\midrule
4  & 16 & 0.021 & $97.9\%$ \\
6  & 24 & 0.028 & $97.2\%$ \\
8  & 32 & 0.034 & $96.6\%$ \\
10 & 40 & 0.041 & $95.9\%$ \\
12 & 48 & 0.052 & $94.8\%$ \\
\bottomrule
\end{tabular}
\end{table}

The TVD increases with system size, as expected from the accumulation of shot noise over larger output distributions. For $N = 12$ and circuit depth 48, the $5.2\%$ TVD is
consistent with statistical sampling noise at $N_{\mathrm{shots}} = 10{,}000$. These results confirm that the tensor network simulator produces correct ideal-state
outputs, verifying implementation correctness.

\subsection{IBM Quantum Hardware Validation}
\label{sec:qpu_hardware}
To validate the simulation framework against real quantum hardware, we execute a suite of
circuits on the IBM \texttt{ibm\_fez} superconducting QPU (156 qubits, Heron r2 processor)
via the Qiskit IBM Runtime service, using $N_{\mathrm{shots}} = 8{,}192$ shots per circuit.
Table~\ref{tab:qpu} reports the fidelity and TVD for four circuit classes.

\begin{table}[htbp]
\centering
\caption{Validation of tensor network simulation against the IBM \texttt{ibm\_fez}
superconducting QPU (156 qubits). Fidelity is defined as the probability of measuring the
target bitstring(s); TVD is the total variation distance between QPU and ideal distributions.
$N_{\mathrm{shots}} = 8{,}192$ for all circuits.}
\label{tab:qpu}
\begin{tabular}{@{}lcccc@{}}
\toprule
Circuit & Qubits & Fidelity & TVD & Notes \\
\midrule
Bell state                   & 2 & 0.940 & 0.060 & $\{|00\rangle{+}|11\rangle\}/\sqrt{2}$ \\
GHZ-4                        & 4 & 0.856 & 0.145 & $\{|0000\rangle{+}|1111\rangle\}/\sqrt{2}$ \\
Variational ansatz (8 params) & 4 & ---   & 0.055 & 2-layer $R_y$-CNOT ansatz \\
Identity (error test)         & 4 & 0.929 & ---   & $U U^\dagger$ circuit \\
\bottomrule
\end{tabular}
\end{table}

The Bell state fidelity of 0.940 and GHZ-4 fidelity of 0.856 are consistent with
the two-qubit gate error rates ($\sim$0.5--1\%) reported for the \texttt{ibm\_fez}
backend. The identity circuit, which constructs and then reverses a GHZ-4 circuit
($H{\cdot}\mathrm{CNOT}^3{\cdot}\mathrm{CNOT}^3{\cdot}H$), achieves a $|0000\rangle$
fidelity of 0.929, providing a direct measurement of cumulative gate error.

For the 4-qubit variational ansatz, the QPU output distribution agrees with the ideal
simulation to within TVD = 0.055. Parameter-shift gradients computed on the QPU agree
with simulator gradients to within a mean absolute error of 0.012, confirming that
QPU noise does not corrupt gradient estimation at the precision level relevant
for variational optimization.

These QPU results complement the SV1 cross-validation above: SV1 verifies
implementation correctness (both are noiseless simulators), while the IBM
QPU data confirms that the simulated distributions are consistent with
real-hardware measurement outcomes at the expected noise level.

\subsubsection{Cross-Vendor QPU Comparison}
\label{sec:cross_vendor_qpu}
To verify that the simulation framework produces hardware-agnostic results, we
execute the same circuit suite on four QPU backends spanning two qubit
technologies: three superconducting processors---IBM \texttt{ibm\_fez} (156~qubits,
Heron~r2), Rigetti Ankaa-3 (82~qubits), and IQM Garnet (20~qubits)---and one
trapped-ion processor, IonQ Forte-1 (36~qubits). Table~\ref{tab:cross_vendor}
reports the results.

\begin{table}[htbp]
\centering
\caption{Cross-vendor QPU validation. Bell and GHZ-4 fidelities, variational
TVD, and parameter-shift gradient metrics across four quantum processors from
independent vendors. IBM: $N_{\mathrm{shots}}=4{,}096$; Rigetti and IQM:
$N_{\mathrm{shots}}=4{,}096$; IonQ: $N_{\mathrm{shots}}=100$.}
\label{tab:cross_vendor}
\begin{tabular}{@{}llccccc@{}}
\toprule
\textbf{Vendor} & \textbf{Backend} & \textbf{Technology} & \textbf{Bell~$F$} & \textbf{GHZ-4~$F$} & \textbf{Var.\ TVD} & \textbf{Grad.\ MAE} \\
\midrule
IBM     & ibm\_fez   & Superconducting & 0.943  & 0.874  & --- & 0.026 \\
Rigetti & Ankaa-3    & Superconducting & 0.933  & 0.763  & 0.124 & 0.012 \\
IQM     & Garnet     & Superconducting & 0.950  & 0.870  & 0.063 & 0.010 \\
IonQ    & Forte-1    & Trapped-ion     & 0.990  & 0.970  & 0.163 & ---   \\
\bottomrule
\end{tabular}
\end{table}

The trapped-ion IonQ Forte-1 achieves the highest Bell and GHZ-4 fidelities
(0.990 and 0.970, respectively), consistent with the lower two-qubit gate
error rates typical of trapped-ion systems. Among the superconducting backends,
IQM Garnet and IBM \texttt{ibm\_fez} show comparable fidelities ($F_{\mathrm{GHZ\text{-}4}}
\approx 0.87$), while Rigetti Ankaa-3 exhibits higher decoherence on the GHZ-4
circuit ($F = 0.763$). Parameter-shift gradients on all superconducting backends
agree with simulator predictions to within MAE $\leq 0.026$, confirming that
our framework's gradient estimation is robust across hardware platforms.

\section{Discussion}
\label{sec:discussion}
The adaptive bond dimension framework presented in this paper addresses a practical limitation
of existing tensor network simulation tools. By replacing user-specified fixed bond dimensions
with closed-loop entropy-feedback control, the framework automates a tedious aspect of MPS
simulation configuration while improving computational efficiency through per-bond resource
allocation and GPU acceleration.

\subsection{Advantages and Scope}

The primary advantage of the adaptive approach is the elimination of manual bond dimension
tuning. In production tensor network workflows, practitioners typically run preliminary
simulations at several $\chi$ values to determine the minimum bond dimension that achieves the
desired accuracy. This trial-and-error process consumes both human time and computational
resources. The PID controller automates this determination, converging to an appropriate $\chi$
profile within approximately 8 sweeps.

The per-bond granularity provides a secondary advantage: the total computational cost is
reduced because bonds with low entanglement use smaller bond dimensions. For the Heisenberg
chain benchmarks, this yields a $2.7\times$ wall-time reduction at $N = 100$. The benefit is
expected to be larger for systems with more inhomogeneous entanglement profiles, such as
impurity models, chains with disorder, or systems near quantum critical points where the
entanglement varies strongly with position.

\subsection{Counterarguments and Limitations}

Several counterarguments and limitations should be acknowledged. First, the PID controller
introduces three additional hyperparameters ($K_p$, $K_i$, $K_d$) that must be tuned. While the
Ziegler-Nichols method provides a systematic tuning procedure, the gains depend on the model
Hamiltonian and system size. The values $K_p = 2.0$, $K_i = 0.1$, $K_d = 0.5$ work well for
the Heisenberg model but may require re-tuning for systems with different entanglement scaling
(for example, critical systems with logarithmic violations of the area law, or frustrated
magnets). Empirically, the controller is robust to gain perturbations: scaling all three gains
by factors $0.5\times$, $0.75\times$, $1.25\times$, and $1.50\times$ on the 20-site Heisenberg
chain yields the same ground-state energy ($E/N = -1.73645$) and the same sweep count (5) across
all settings. On the critical Ising chain, the same gain perturbations ($0.5\times$ through
$1.5\times$) likewise converge to the same $E/N = -1.25539$ in 4 sweeps, confirming gain
robustness across Hamiltonian classes.

Second, the adaptive framework adds computational overhead from the entropy computation, EMA
filtering, and PID update at each bond. This overhead is $\mathcal{O}(N \chi)$ per sweep (for
the entropy computation from $\chi$ singular values at each of $N-1$ bonds), compared with the
$\mathcal{O}(N d \chi^3)$ cost of the SVD operations. The overhead is therefore negligible for
$\chi \gg d$, which is the regime where adaptive bond dimension management provides the greatest
benefit. For very small $\chi$ (below approximately 16), the adaptive framework provides little
advantage because the bond dimension has insufficient dynamic range for meaningful adaptation.

Third, the GPU acceleration is effective only for sufficiently large bond dimensions
($\chi \gtrsim 64$). For small systems or low-entanglement states where $\chi$ remains below
the CPU/GPU crossover, the CPU-only implementation is preferable. The hybrid strategy (GPU for
$\chi \geq 64$, CPU otherwise) addresses this limitation but introduces a discontinuity in
computational performance at the threshold.

Fourth, the predictive scheduler of Section~\ref{sec:predictive} assumes approximately linear
entropy growth over short time horizons. For systems undergoing abrupt entanglement transitions
(such as quench dynamics across a quantum phase transition), the linear extrapolation may
significantly underestimate future entropy requirements, causing the PID controller to lag behind
the entanglement growth. In such cases, a more sophisticated predictor or a larger safety margin
$\gamma$ may be necessary.

Fifth, the framework has been tested primarily on 1D systems. Extension to two-dimensional
projected entangled pair states (PEPS) or tree tensor networks~\cite{Shi2006} would require
adapting the per-bond PID controllers to the different TN geometry and entanglement scaling.
The bond dimension management concept is general, but the specific PID tuning and entropy-$\chi$
relationship would differ for non-MPS tensor networks.

Sixth, the stability analysis in Section~\ref{sec:pid} generalizes the linearization to account
for gain uncertainty across operating regimes: the loop gain $g(\chi^*)$ varies from
$g_{\mathrm{sub}} \ll 1/\chi^*$ at sub-saturated bonds (gapped Hamiltonians with exponentially
decaying singular values) to $g_{\mathrm{sat}} = 1/\chi^*$ at saturation and
$g_{\mathrm{crit}} \leq 1/(\chi^* \ln \chi^*)$ at critical points. The Jury stability
criterion confirms all poles lie within the unit circle for $g \in (0, 3/\chi^*)$, covering
the full sub-saturated and saturated regimes ($\chi^* \geq 8$) and the critical regime
($\chi^* \geq 16$). The $\pm 50\%$ PID gain sweep (Table~\ref{tab:ablation}) and convergence
across gapped/critical Hamiltonians (Table~\ref{tab:hamiltonians}) provide empirical
confirmation. A full Lyapunov analysis incorporating clamp non-linearity and state-dependent
gain trajectories remains open for future work.

Seventh, the ablation study of Table~\ref{tab:ablation} isolates the PID controller's
contribution from the overall benchmark improvement: on the 20-site Heisenberg chain, PID-controlled
adaptive allocation provides a $5.5\times$ speedup over fixed-$\chi$, while threshold-based
truncation achieves only $4.3\times$. The residual $1.3\times$ gap between PID and threshold
quantifies the value of derivative and integral control action. However, a full factorial ablation
combining GPU/CPU $\times$ adaptive/fixed is deferred to a follow-up study on larger systems
where GPU effects are measurable ($\chi \geq 64$).

Eighth, we validate the predictive scheduler in isolation by comparing
reactive-only control ($\beta = 0$) against predictive control ($\beta = 0.4$, $0.8$) on the
20-site Heisenberg and critical Ising chains. For both Hamiltonians, all three $\beta$ values
converge to the same ground-state energy in the same number of sweeps ($5$ for Heisenberg,
$4$ for Ising critical). The wall times are statistically indistinguishable ($3.59$--$4.12$\,s),
indicating that the predictor does not degrade performance but provides minimal benefit for
systems where the entanglement profile is stable between sweeps. The predictor is expected to
show larger benefit in time-dependent simulations where entanglement grows monotonically.

\subsection{Comparison with Alternative Approaches}

An alternative to PID-based control is a simple threshold-based truncation, where singular values
below $\epsilon_{\mathrm{trunc}}$ are discarded regardless of how many remain. This approach,
commonly implemented in existing libraries, does provide some adaptivity in the effective bond
dimension. However, it does not incorporate feedback dynamics (it is purely reactive with no
integral or derivative action), cannot anticipate entanglement growth, and does not benefit from
the smoothing and stability properties of the EMA-PID combination. In our benchmarks,
threshold-based truncation with $\epsilon_{\mathrm{trunc}} = 10^{-10}$ achieves comparable
accuracy to the adaptive framework but $1.4\times$ longer wall time for the 100-site Heisenberg
chain, because it tends to over-allocate bond dimension at boundary bonds during early sweeps.

\section{Conclusion}
\label{sec:conclusion}
This paper has presented an adaptive tensor network simulation framework that uses entropy-feedback
PID control to dynamically manage bond dimensions in Matrix Product State calculations. The
framework replaces the conventional fixed bond dimension strategy with a closed-loop control system
that monitors von Neumann entanglement entropy at each bond, smooths the measurement with an
exponential moving average filter, and applies PID control to adjust the bond dimension in response
to local entanglement requirements.

The per-bond granularity of the allocation concentrates computational resources at
high-entanglement bonds while reducing the bond dimension at low-entanglement boundaries and
chain edges. A predictive scheduling module extrapolates entropy trends to reduce the controller
lag during rapid entanglement growth.

GPU-accelerated SVD via CuPy and the cuSOLVER backend provides speedups of $4.1\times$ at
$\chi = 256$, $4.4\times$ at $\chi = 512$, and $7.1\times$ at $\chi = 2048$ relative to CPU-based NumPy, with the benefit
increasing for larger bond dimensions (measured on NVIDIA A100-SXM4-40GB, CuPy 13.4.1, CUDA 12.8). Integration with the DMRG algorithm yields ground-state
energies for the Heisenberg chain converging to the Bethe ansatz value at $\chi = 128$, with a
$2.7\times$ wall-time reduction compared to fixed-$\chi$ DMRG at 100 sites.

Validation against the AWS Braket SV1 statevector simulator confirms simulation-implementation agreement within
$2$--$5\%$ total variation distance for systems up to 12 qubits. Real-hardware validation on the
IBM \texttt{ibm\_fez} QPU (156 qubits, Heron r2 processor) yields a Bell-state fidelity of 0.940,
a 4-qubit GHZ fidelity of 0.856, and variational ansatz TVD of 0.055, confirming consistency
between simulation and superconducting QPU outputs. The framework is implemented in
Python with NumPy, SciPy, and CuPy backends, providing compatibility with the scientific Python
ecosystem.

Controlled ablation on the 20-site Heisenberg chain shows that PID-controlled adaptive allocation
contributes a $5.5\times$ speedup over fixed-$\chi$ (compared to $4.3\times$ for threshold-only),
and the controller is robust to $\pm 50\%$ gain perturbations across both the Heisenberg and
critical Ising models. Multi-Hamiltonian benchmarks across four distinct spin chains confirm
speedups of $2.4$--$5.1\times$ with energy accuracy losses below $5\times10^{-5}$ per site.

Future work will extend the adaptive framework to time-dependent
simulations~\cite{Paeckel2019,Daley2004}, where the entanglement growth during unitary evolution
poses an additional challenge for bond dimension management.  Application to two-dimensional tensor
network geometries~\cite{Stoudenmire2012,Verstraete2008} and integration with established
libraries such as ITensor~\cite{Fishman2022} and TeNPy~\cite{Hauschild2018} are also planned.

\section*{Data Availability}
The benchmark data and analysis scripts used in this work will be provided upon reasonable request. The framework requires Python~3.10+, NumPy, SciPy, and CuPy.

\section*{Acknowledgements}
The authors acknowledge computational resources of the Intelligent Robotics and Rebooting Computing Chip Design (INTRINSIC) Laboratory, Centre for SeNSE, Indian Institute of Technology Delhi, IM00002G\_RB\_SG IoE Fund Grant (NFSG), Indian Institute of Technology Delhi.

\section*{Author Contributions (CRediT)}

\textbf{Santhosh Sivasubramani}: Conceptualization, Methodology, Software (architecture
and core implementation), Investigation, Validation, Writing -- original draft, Writing --
review \& editing, Supervision, Project administration, Funding acquisition.
\textbf{Harshni Kumaresan}: Data curation, Formal analysis, Visualization, Writing --
review \& editing.
\textbf{Gayathri Muruganantham}: Data curation, Formal analysis, Visualization, Writing --
review \& editing.
\textbf{Lakshmi Rajendran}: Data curation, Formal analysis, Visualization, Writing --
review \& editing.
\medskip \noindent All authors have reviewed and agreed to the published version of the manuscript.

\section*{Conflict of Interest}
Authors declare no competing financial or non-financial interests.

\bibliographystyle{unsrtnat}
\bibliography{references}

\end{document}